\documentclass[runningheads]{llncs}
\usepackage[T1]{fontenc}
\usepackage{amsmath,amsfonts,epsfig}
\usepackage{makecell}
\usepackage{graphicx}
\usepackage{pdfpages}
\usepackage{subfigure}
\usepackage{wrapfig}
\usepackage{tabularx}
\usepackage{appendix}
\usepackage[misc]{ifsym}
\usepackage{amssymb}
\usepackage[ruled,vlined]{algorithm2e}
\begin{document}
\title{Mastering Complex Coordination through Attention-based Dynamic Graph}
%
%
\author{Guangchong Zhou\inst{1,2} \and Zhiwei Xu\inst{1,2} \and Zeren Zhang\inst{1,2} \and Guoliang Fan\inst{1}\Letter}
\authorrunning{G. Zhou et al.}
%
\institute{Institute of Automation, Chinese Academy of Sciences \and School of Artificial Intelligence, University of Chinese Academy of Sciences\\Beijing, China\\
\email{$\{$zhouguangchong2021, xuzhiwei2019, zhangzeren2021, guoliang.fan$\}$@ia.ac.cn}}
\toctitle{Mastering Complex Coordination through Attention-based Dynamic Graph}
\tocauthor{Guangchong Zhou, Zhiwei Xu, Zeren Zhang, Guoliang Fan}
\maketitle              
\begin{abstract}
The coordination between agents in multi-agent systems has become a popular topic in many fields. To catch the inner relationship between agents, the graph structure is combined with existing methods and improves the results. But in large-scale tasks with numerous agents, an overly complex graph would lead to a boost in computational cost and a decline in performance. Here we present DAGMIX, a novel graph-based value factorization method. Instead of a complete graph, DAGMIX generates a dynamic graph at each time step during training, on which it realizes a more interpretable and effective combining process through the attention mechanism. Experiments show that DAGMIX significantly outperforms previous SOTA methods in large-scale scenarios, as well as achieving promising results on other tasks.

\end{abstract}
\section{Introduction}
Multi-agent systems (MAS) have become a popular research topic in the last few years, due to their rich application scenarios like auto-driving~\cite{shamsoshoara2019distributed}, cluster control~\cite{xu2022autonomous}, and game AI~\cite{vinyals2019grandmaster,berner2019dota,ye2020towards}. Communication constraints exist in many MAS settings, meaning only local information is available for agents' decision-making. A lot of research discussed multi-agent reinforcement learning (MARL), which combines MAS and reinforcement learning techniques. The simplest way is to employ single-agent reinforcement learning on each agent directly, as the \textit{independent Q-learning} (IQL)~\cite{tampuu2017multiagent} does. But this approach may not converge as each agent faces an unstable environment caused by other agents' learning and exploration. Alternatively, we can learn decentralized policies in a centralized fashion, known as the \textit{centralized training and decentralized execution} (CTDE)~\cite{lowe2017multi} paradigm, which allows sharing of individual observations and global information during training but limits the agents to receive only local information while executing. 

Focusing on the cooperative tasks, the agents suffer from the credit assignment problem. If all agents share a joint value function, it would be hard to measure each agent's contribution to the global value, and one agent may mistakenly consider others' credits as its own. Value factorization methods decompose the global value function as a mixture of each agent's individual value function, thus solving this problem from the principle. Some popular algorithms such as VDN~\cite{sunehag2017value}, QMIX~\cite{rashid2018qmix}, and QTRAN~\cite{DBLP:journals/corr/abs-1905-05408} have demonstrated impressive performance in a range of cooperative tasks.

\begin{wrapfigure}{H}{0.45\textwidth}
    \begin{center}
        \vspace{-1.2cm}
        \includegraphics[width=0.4\textwidth, height=0.25\textwidth]{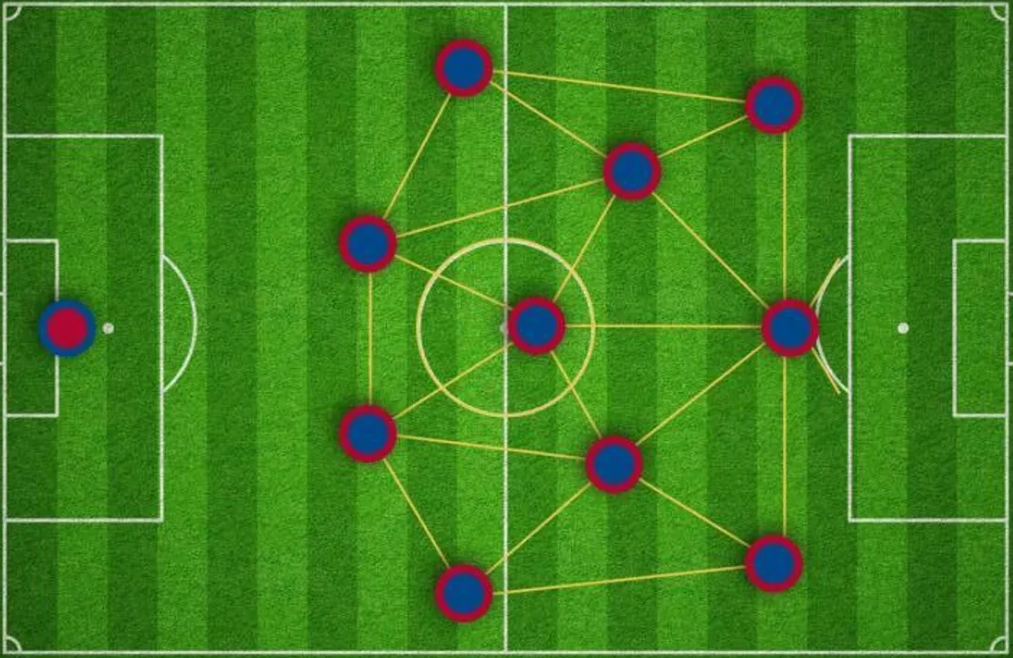}
        \caption{Topology of football players.}
        \vspace{-1cm}
        \label{fig:football}
    \end{center}
\end{wrapfigure}

The above-mentioned algorithms focus on constructing the mixing function mathematically but ignore the physical topology of the agents. For example, players in a football team could form different formations like '4-3-3' or '4-3-1-2' (shown in Figure~\ref{fig:football}), and a player has stronger links with the nearer teammates. An intuitive idea is to use graph neural networks (GNN) to extract the information contained in the structure. \textit{Multi-agent graph network} (MAGNet)~\cite{malysheva2019magnet} generates a real-time graph to guide the agents' decisions based on historical and current global information. In QGNN~\cite{kortvelesy2022qgnn}, the model encodes every agent's trajectory by a gated recurrent unit (GRU), which is followed by a GNN to produce a local Q-value of each agent. These methods do make improvements on original algorithms, but they violate the communication constraints, making it hard to extend to more general scenarios. Xu et al. modify QMIX and propose \textit{multi-graph attention network} (MGAN)~\cite{xu2021learning}, which constructs all agents as a graph and utilizes multiple graph convolutional networks (GCN~\cite{kipf2016semi}) to jointly approximate the global value. Similar but not the same, \textit{Deep Implicit Coordination Graph} (DICG)~\cite{li2020deep} processes individual Q-values through a sequence of GCNs instead of parallel ones. These algorithms fulfill CTDE and significantly outperform the benchmarks. However, these methods prefer fully connected to sparse when building the graph. As the number of agents grows, the complexity of a complete graph increases by square, leading to huge computational costs and declines in the results.

In this paper, we proposed a cooperative multi-agent reinforcement learning algorithm called \emph{Dynamic Attention-based Graph MIX} (DAGMIX). DAGMIX presents a more reasonable and intuitive way of value mixing to estimate $Q_{tot}$ more accurately, thus providing better guidance for the learning of agents. Specifically, DAGMIX generates a partially connected graph rather than fully connected according to the agents' attention on each other at every time step during training. Later we perform graph attention on this dynamic graph to help integrate individual Q-values. Like other value factorization methods, DAGMIX perfectly fulfills the CTDE paradigm, and it also meets the monotonicity assumption, which ensures the consistency of global optimal and local optimal policy. Experiments on StarCraft multi-agent challenge (SMAC)~\cite{samvelyan2019starcraft} show that our work is comparable to the baseline algorithms. DAGMIX outperforms previous SOTA methods significantly in some tasks, especially those with numerous agents and asymmetric environmental settings which are considered as super-hard scenarios.

\section{Background}
\subsection{Dec-POMDP}
A fully cooperative multi-agent environment is usually modeled as a decentralized partially observable Markov decision process (Dec-POMDP)~\cite{oliehoek2016concise}, consisting of a tuple $G = \langle\mathcal{S},\boldsymbol{\mathcal{U}}, \mathcal{P}, \mathcal{Z}, r, \mathcal{O}, n, \gamma\rangle$. At each time-step, $s \in \mathcal{S}$ is the current global state of the environment, but each agent $a \in \mathcal{A} := \left \{1, ..., n\right \}$ receives only a unique local observation $z_a \in \mathcal{Z}$ produced by the observation function $\mathcal{O}(s, a): \mathcal{S} \times \mathcal{A} \rightarrow \mathcal{Z}$. Then every agent $a$ will choose an action $u_a \in \mathcal{U}$, and all individual actions form the joint action $\boldsymbol{u}=[u_1, ...,.u_n] \in \boldsymbol{\mathcal{U}} \equiv \mathcal{U}^n$. After the interaction between the joint action $\boldsymbol{u}$ and current state $s$, the environment changes to state $s^{\prime}$ according to state transition function $\mathcal{P}(s'\|s, \boldsymbol{u}): \mathcal{S} \times \mathcal{U} \times \mathcal{S} \rightarrow [0, 1]$. All the agents in Dec-POMDP share the same global reward function $r(s, \boldsymbol{u}): \mathcal{S} \times \boldsymbol{\mathcal{U}} \rightarrow \mathbb{R}$. $\gamma \in [0, 1)$ is the discount factor. 

In Dec-POMDP, each agent $a$ chooses action based on its own action-observation history $\tau_a \in T \equiv(\mathcal{Z}\times \mathcal{U})$, thus the policy of each agent $a$ can be written as $\pi_a (u_a | \tau_a) : T \times \mathcal{U} \rightarrow \left[0, 1\right]$. The joint action-value function can be computed by the following equation: $Q^{\boldsymbol{\pi}}(s_t, \boldsymbol{u}_t) = \mathbb{E}_{s_{t+1:\infty},\boldsymbol{u}_{t+1:\infty}}\left[R_t|s_t,\boldsymbol{u}_t\right]$, where $\boldsymbol{\pi}$ is the joint policy of all agents. The goal is to maximize the discounted return $R^t = \sum_{l=0}^\infty \gamma^lr_{t+l}$. 

\subsection{Value Factorization Methods}
Credit assignment is a key problem in cooperative MARL problems. If all agents share a joint value function, it would be hard for a single agent to tell how much it contributes to global utilization. Without such feedback, learning is easy to fail. 

In value factorization methods, every agent has its own value function for decision-making. The joint value function is regarded as an integration of individual ones. To ensure that the optimal action of each agent is consistent with the global optimal joint action, all value decomposition methods comply with the \textit{Individual Global Max} (IGM)~\cite{rashid2018qmix} conditions described below:
\begin{equation*}
    \arg \max _{\boldsymbol{u}} Q_{\mathrm{tot}}(\boldsymbol{\tau}, \boldsymbol{u})=\left(\begin{array}{c}
    \arg \max _{u_{1}} Q_{1}\left(\tau_{1}, u_{1}\right) \\
    \vdots \\
    \arg \max _{u_{n}} Q_{n}\left(\tau_{n}, u_{n}\right)
    \end{array}\right),
\end{equation*}
where $Q_{tot}=f(Q_1, ..., Q_n)$, $Q_1, ..., Q_n$ denote the individual Q-values, and $f$ is the mixing function.

For example, VDN assumes the function $f$ is a plus operation, while QMIX makes monotonicity constraints to meet the IGM conditions as below:
\begin{align*}
   \left(\textrm{VDN}\right) \qquad Q_{tot}(s,u_a) &= \sum_{a=1}^n Q_a(s,u_a).\\
    \left(\textrm{QMIX}\right) \qquad \frac{\partial Q_{tot}(\boldsymbol{\tau}, \boldsymbol{u})}{\partial Q_{a}\left(\tau_{a}, u_{a}\right)} &\geq 0, \quad \forall a \in \{1,\dots,n\}.
\end{align*}

\subsection{Self Attention}
Consider a sequence of vectors $\boldsymbol{\alpha}=\left(\mathbf{a}_1, ..., \mathbf{a}_n\right)$, the motivation of self-attention is to catch the relevance between each pair of vectors~\cite{vaswani2017attention}. An attention function can be described as mapping a query ($\mathbf{q}$) and a set of key-value ($\mathbf{k}, \mathbf{v}$) pairs to an output ($\mathbf{o}$), where the query, keys, and values are all translated from the original vectors. The output is derived as below: 
\begin{equation*}
    w_{ij}=\textrm{softmax}\left(s\left(\mathbf{q}^i, \mathbf{k}^j\right)\right)
    =\frac{\exp\left(s\left(\mathbf{q}^i, \mathbf{k}^j\right)\right)}{\sum_t \exp\left(s\left(\mathbf{q}^i, \mathbf{k}^t\right)\right)},
\end{equation*}
\begin{equation*}
    \mathbf{o}_i=\sum_j w_{ij}\mathbf{v}_j,
\end{equation*}

where $s\left(\cdot, \cdot\right)$ is a user-defined function to measure similarity, usually dot-product. In practice, a multi-head implementation is usually employed to enable the model to calculate attention from multiple perspectives.

\subsection{Graph Neural Network}
\textit{Graph Neural Networks} (GNNs) are a class of deep learning methods designed to directly perform inference on data described by graphs and easily deal with node-level, edge-level, and graph-level tasks.

The main idea of GNN is to aggregate the node's own information and its neighbors' information together using a neural network. Given a graph denoted as $\boldsymbol{G}=\left(\boldsymbol{V}, \boldsymbol{E}\right)$ and consider a $K$-layer GNN structure, the operation of the $k$-th layer can be formalized as below~\cite{hamilton2017inductive}:
\begin{equation*}
    \mathbf{h}_v^k=\sigma \left(\left[\mathbf{W}_k \cdot \textrm{AGG} \left(\left\{\mathbf{h}_u^{k-1}, \forall u \in \mathcal{N}(v) \right\} \right), \mathbf{B}_k \mathbf{h}_v^{k-1} \right] \right),
\end{equation*}
where $\mathbf{h_v^k}$ refers to the value of node $v$ at $k$-th layer, $\mathcal{N}$ is the neighborhood function to get the neighbor nodes, and $\textrm{AGG}\left(\;\hdots\; \right)$ is the generalized aggregator which can be a function such as Mean, Pooling, LSTM and so on.

Unlike other deep networks, GNN could not improve its performance by simply deepening the model, since too many layers in GNN usually lead to severe over-smoothing or over-squashing. Besides, the computational complexity of GNN is closely related to the amount of nodes and edges. Fortunately, DAGMIX tries to generate an optimized graph structure, making it suitable for applying GNN.

\section{DAGMIX}
\begin{figure*}[t]
    \centering
    \includegraphics[width = 0.99\textwidth]{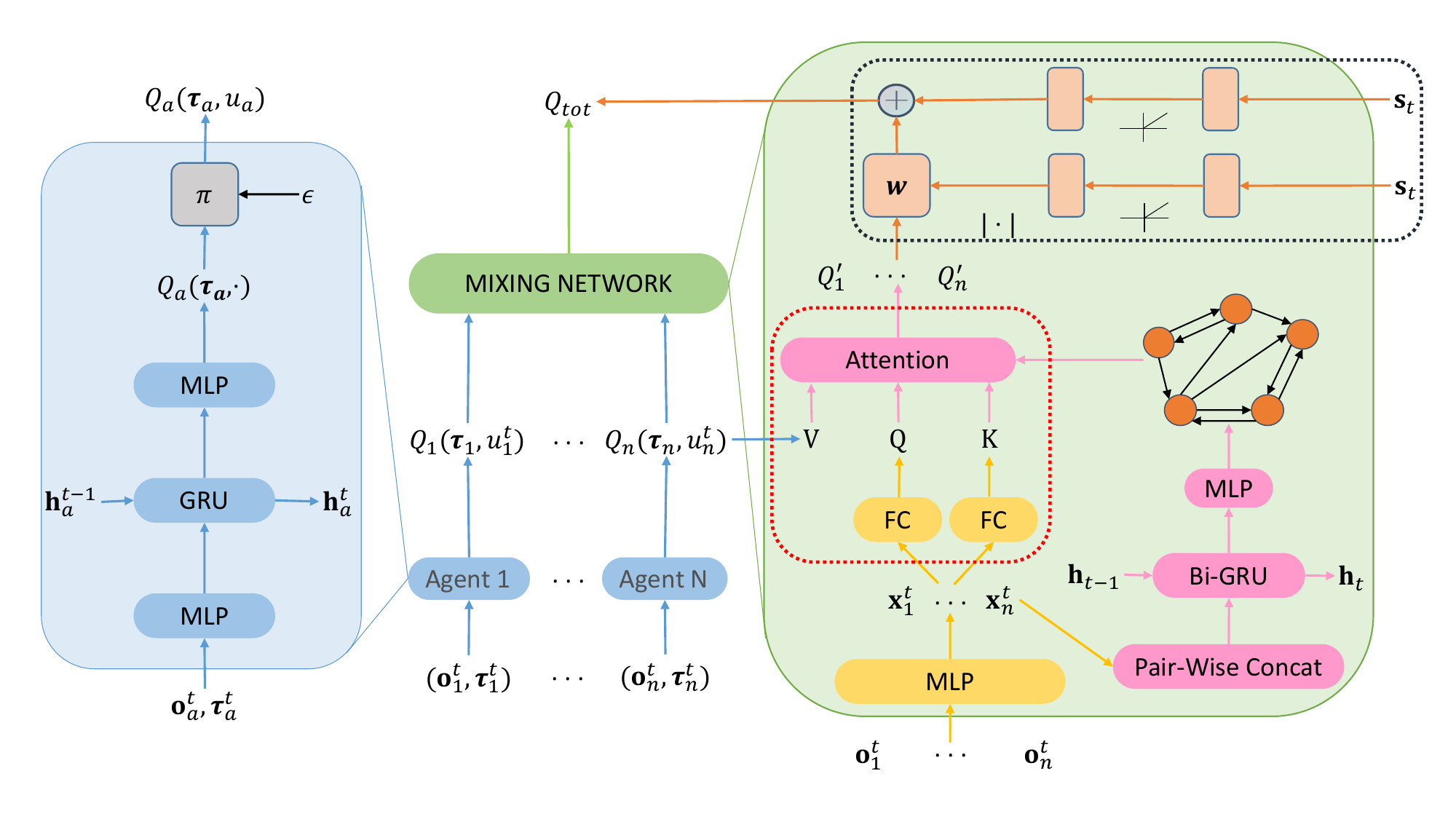}
    \caption{The overall framework of DAGMIX. The individual Q network is shown in the blue box, and the mixing network is shown in the green box.}
    \label{fig:DAGMIX}
\end{figure*}

In this section, we'd like to introduce a new method called DAGMIX. Compared to other related studies, DAGMIX enables a dynamic graph in the training process of cooperative MARL. It fulfills the CTDE paradigm and IGM conditions perfectly and provides a more precise and explainable estimation of all agents' joint action, thus reaching a better performance than previous methods. The overall architecture is shown in Figure~\ref{fig:DAGMIX}.

\subsection{Dynamic Graph Generation}
Each agent corresponds to a node in the graph. Instead of relying on a fully connected graph, DAGMIX employs a dynamic graph that generates a real-time structure at each time step. Inspired by G2ANet~\cite{liu2020multi}, we adopt the hard-attention mechanism. It computes the attention weights between all the nodes and then sets the weights to 0 or 1 representing if there are connections between each pair of nodes in the dynamic graph. Nonetheless, DAGMIX removes the communication between agents in G2ANet and enables fully decentralized decision-making.

Assume that we take out an episode from the replay buffer. First, we encode the observation $o_a^t$ of each agent $a$ at each time-step $t$ as the embedding $\mathbf{x}_a^t$. Then we perform a pair-wise concatenation on all the embedding vectors, which yields a matrix $\mathbf{x}_{n\times n}^t$ where $\mathbf{x}_{ij}^t=\left(\mathbf{x}_i^t, \mathbf{x}_j^t \right), \quad i,j=1,...,n$. This matrix $\mathbf{x}^t$ can be regarded as a batch of sequential data since it has a batch size of $n$ corresponding to $n$ agents, and every agent $a$ has a sequence from $(\mathbf{x}_a^t, \mathbf{x}_1^t)$ to $(\mathbf{x}_a^t, \mathbf{x}_n^t)$. We could naturally think of using a sequential model like GRU to process every sequence and calculate the attention between agents.

Notably, the output of traditional GRU only depends on the current and previous inputs but ignores the subsequent ones. Therefore, the agent at the front of the sequence has a greater influence on the output than the agent at the end, making the order of the agents really crucial. Such a mechanism is apparently unjustified for all agents, so we adopt a bi-directional GRU (Bi-GRU) to fix it. The calculation on the concatenation $\left(\mathbf{x}_i^t, \mathbf{x}_j^t \right)$ is formalized by Equation~(\ref{eq:Bi-GRU}), where $i,j=1,...,n$ and $f\left(\cdot\right)$ is a fully connected layer to embed the output of GRU.

\begin{equation}
    \mathbf{a}_{i,j}=f\left(\textrm{Bi-GRU}\left(\mathbf{x}_i, \mathbf{x}_j \right)\right).
    \label{eq:Bi-GRU}
\end{equation}

Then we need to decide if there's a link between agent $i$ and $j$. One approach is to sample on 0 and 1 based on $\mathbf{a}_{i, j}$, but simply sampling is not differentiable and makes the graph generation module untrainable. To maintain the gradient continuity, DAGMIX adopts gumbel-softmax~\cite{jang2016categorical} as follows:

\begin{equation}
    \textrm{gum}\left(\mathbf{x}\right) = \textrm{softmax}\left(\left(\mathbf{x}+\log \left(\lambda e^{-\lambda \mathbf{x}}\right)\right) / \tau\right),
    \label{eq:gumbel-softmax}
\end{equation}

where $\textrm{gum}\left(\cdot\right)$ means the gumbel-softmax function, $\lambda$ is the hyperparameter of the exponential distribution, and $\tau$ is the hyperparameter referring to the temperature. As the temperature decreases, the output of Gumbel-softmax gets closer to one-hot. In DAGMIX, Gumbel-softmax returns a vector of length 2.

\begin{equation}
    \left(\;\cdot\;,\mathbf{A}_{i,j}\right)^{\mathrm{T}}=\textrm{gum}\left(f\left(\textrm{Bi-GRU}\left(\mathbf{x}_i, \mathbf{x}_j \right)\right)\right).
    \label{eq:AdjMatrix}
\end{equation}

Combining Equation~(\ref{eq:gumbel-softmax}) and Equation~(\ref{eq:Bi-GRU}) we get Equation~(\ref{eq:AdjMatrix}). $\mathbf{A}$ is the adjacency matrix whose element $\mathbf{A}_{i,j}$ is the second value in the output of Gumbel-softmax, indicating whether there's an edge from agent $j$ to agent $i$. Finally, we get a graph structure similar to the one illustrated on the right side of Figure~\ref{fig:DAGMIX}. It is characterized by a sparse connectivity pattern that exhibits an expansion in sparsity as the graph scales up.

\subsection{Value Mixing Network}
As we've already got a graph structure~(shown in the right of Figure~\ref{fig:DAGMIX}), we design an attention-based value integration on the dynamic graph, which can be viewed as performing self-attention on the original observations. 

More specifically, there are two independent channels to process the observation. On the one hand, the Q network receives current observation as input and outputs an individual Q value estimation, which serves as the value in self-attention. On the other hand, the observation is encoded with an MLP in the mixing network, which is followed by two fully connected layers representing $\mathbf{W}_q$ and $\mathbf{W}_k$. The query and key of agent $a$ in self attention are derived through $\boldsymbol{q}_a=\mathbf{W}_q \mathbf{x}_a$ and $\boldsymbol{k}_a=\mathbf{W}_k \mathbf{x}_a$ respectively.

Besides, restricted by the dynamic graph, we only calculate the attention scores in agent $a$'s neighborhood $\mathcal{N}\left(a\right) = \{i\in \mathcal{A}\vert \mathbf{A}_{a,i}=1\}$, which can be regarded as a mask operation based on the adjacency matrix $\mathbf{A}$. The calculation process in matrix form is shown below:
\begin{align}
&\boldsymbol{Q}=\boldsymbol{X}\mathbf{W}_q, \nonumber \quad
\boldsymbol{K}=\boldsymbol{X}\mathbf{W}_k, \nonumber \quad
\boldsymbol{V}=\left(Q_1, ..., Q_n\right)^{\mathrm{T}}, \nonumber \\
&\left(Q_1^{'}, ..., Q_n^{'}\right)^{\mathrm{T}} = \textrm{softmax}\left(\textrm{Mask}\left(\frac{\boldsymbol{Q}\boldsymbol{K}^{\mathrm{T}}}{\sqrt{d_k}}\right)\right)\boldsymbol{V},
\label{eq:Attention}
\end{align}

where $\boldsymbol{X}=\left(\mathbf{x}_1, ..., \mathbf{x}_n\right)^{\mathrm{T}}$, and $d_k$ is the hidden dimension of $\mathbf{W}_k$. According to Equation~(\ref{eq:Attention}), DAGMIX blends the original Q-values of all agents and transforms them into $\left(Q_1^{'}, ..., Q_n^{'}\right)$. 

In the last step, we need to combine $\left(Q_1^{'}, ..., Q_n^{'}\right)$ into a global $Q_{tot}$. As the dotted black box in Figure~\ref{fig:DAGMIX} shows, we adopt a QMIX-style mixing module, making DAGMIX benefit from the global information. The hypernetworks~\cite{ha2016hypernetworks} take in global state $\mathbf{s}_t$ and output the weights $\mathbf{w}$ and bias $b$, thus the $Q_{tot}$ is calculated as:
\begin{equation}
    Q_{tot} = \left(Q_1^{'}, ..., Q_n^{'}\right) \mathbf{w}+b
    \label{eq:qmix}
\end{equation}

The operation of taking absolute values guarantees weights in $\mathbf{w}$ non-negative. And in Equation~\ref{eq:Attention}, the coefficients of any $Q_a$ are also non-negative thanks to the softmax function. Equation~\ref{partial} ensures DAGMIX fits in the IGM assumption perfectly.

\begin{equation}
    \frac{\partial Q_{tot}}{\partial Q_a}=\sum_{i=1}^{n} \frac{\partial Q_{tot}}{\partial Q_i^{'}}\frac{\partial Q_i^{'}}{\partial Q_a} \geq 0, \quad \forall a \in \{1, ..., n\}
    \label{partial}
\end{equation}

It is noteworthy that DAGMIX is not a specific algorithm but a framework. As depicted by the green box in Figure~\ref{fig:DAGMIX}, the part in the red dotted box can be replaced by any kind of GNN~(e.g., GCN, GraphSAGE~\cite{hamilton2017inductive}, etc.), while the mixing module in the black dotted box can be substituted with any value factorization method. DAGMIX enhances the performance of the original algorithm, especially on large-scale problems.

\subsection{Loss Function}
Like other value factorization methods, DAGMIX is trained end-to-end. The loss function is set to TD-error, which is similar to value-based SARL algorithms ~\cite{sutton2018reinforcement}. We denote the parameters of all neural networks as $\theta$ which is optimized by minimizing the following loss function:
\begin{equation}
    \mathcal{L}(\theta) = \left(y_{tot}-Q_{tot}(\boldsymbol{\tau},\boldsymbol{u}|\theta)\right)^2,
    \label{eq:loss}
\end{equation}
where $y_{tot}=r+\gamma \max_{\boldsymbol{u^\prime}}Q_{tot}$ $(\boldsymbol{\tau}^ \prime,\boldsymbol{u}^\prime|\theta^-)$ is the target joint action-value function. $\theta^-$ denotes the parameters of the target network. The training algorithm is displayed in Algorithm~\ref{alg:DAGMIX}.

\begin{algorithm}[t]
\DontPrintSemicolon
\SetAlgoNoLine
Initialize replay buffer $\mathbf{D}$\\
Initialize $[Q_a],Q_{tot}$ with random parameters $\theta$, initialize target parameters $\theta^-=\theta$\\
\While {training}{
\For {$episode \leftarrow 1$ \KwTo $M$}{
Start with initial state $\mathbf{s}^0$ and each agent's observation $\mathbf{o}_a^0=\mathcal{O}\left(\mathbf{s}^0, a\right)$\\
Initialize an empty episode recorder $E$
\For {$t \leftarrow 0$ \KwTo $T$}{
For every agent $a$, with probability $\epsilon$ select action $u_a^t$ randomly\\
Otherwise select $u_a^t=\arg \max_{u_a^t} Q_a\left(\tau_a^t, u_a^t\right)$\\
Take joint action $\boldsymbol{u}^t$, and retrieve next state $\mathbf{s}^{t+1}$, next observations $\boldsymbol{o}^{t+1}$ and reward $r^t$\\
Store transition $\left(\mathbf{s}^t, \boldsymbol{o}^{t}, \boldsymbol{u}^t, r^t, \mathbf{s}^{t+1},\boldsymbol{o}^{t+1}\right)$ in $E$
}
Store episode data $E$ in $\mathbf{D}$
}
Sample a random mini-batch data $\mathbf{B}$ with batch size $N$ from $\mathbf{D}$\\
\For {$t \leftarrow 0$ \KwTo $T-1$}{
Extract transition $\left(\mathbf{s}^t, \boldsymbol{o}^{t}, \boldsymbol{u}^t, r^t, \mathbf{s}^{t+1},\boldsymbol{o}^{t+1}\right)$ from $\mathbf{B}$\\
For every agent $a$, calculate $Q_a\left(\tau_a^t, u_a^t|\theta\right)$\\
Generate the dynamic graph $\mathcal{G}^t$ using $\boldsymbol{o}^{t}$ based on Equation~(\ref{eq:AdjMatrix})\\
According to the structure of $\mathcal{G}^t$, calculate $Q_{tot}\left(\boldsymbol{\tau}^t,\boldsymbol{u}^t|\theta\right)$ based on Equation~(\ref{eq:Attention}) and (\ref{eq:qmix})\\
With target network, calculate $Q_a\left(\tau_a^{t+1}, u_a^{t+1}|\theta^-\right)=\max Q_a\left(\tau_a^{t+1}, \;\cdot\;|\theta^-\right)$\\
Calculate $Q_{tot}\left(\boldsymbol{\tau}^{t+1},\boldsymbol{u}^{t+1}|\theta^-\right)$
}
Update $\theta$ by minimizing the total loss in Equation~(\ref{eq:loss})\\
Update target network parameters $\theta^-=\theta$ periodically
}
\label{alg:DAGMIX}
\caption{DAGMIX}
\end{algorithm}

\section{Experiments}
\subsection{Settings}
SMAC is an environment for research in the field of collaborative multi-agent reinforcement learning (MARL) based on Blizzard's StarCraft II RTS game. It consists of a set of StarCraft II micro scenarios that aim to evaluate how well independent agents are able to learn coordination to solve complex tasks. The version of StarCraft II is 4.6.2(B69232) in our experiments, and it should be noted that results from different client versions are not always comparable. The difficulty of the game AI is set to \textit{very hard}~(7). To conquer a wealth of challenges with varying levels of difficulty in SMAC, algorithms should adapt to different scenarios and perform well both in single-unit control and group coordination. SMAC has become increasingly popular recently for its ability to comprehensively evaluate MARL algorithms.

The detailed information of the challenges used in our experiments is shown in Table~\ref{tab:challenges}. All the selected challenges have a large number of agents to control, and some of them are even asymmetric or heterogeneous, making it extremely hard for MARL algorithms to handle these tasks. 

Our experiment is based on Pymarl~\cite{samvelyan2019starcraft}. To judge the performance of DAGMIX objectively, we adopt several most popular value factorization methods (VDN, QMIX, and QTRAN) as well as  more recent methods including Qatten~\cite{yang2020qatten}, QPLEX~\cite{wang2020qplex}, W-QMIX~\cite{rashid2020weighted}, MAVEN~\cite{mahajan2019maven}, ROMA~\cite{wang2020roma} and RODE~\cite{wang2020rode} as baselines. The hyperparameters of these baseline algorithms are set to the default in Pymarl.

\begin{table}[h]
\small
\centering
\caption{Information of selected challenges.}
\resizebox{.9\linewidth}{!}{
\begin{tabular}{lccc}
\hline
Challenge&Ally Units&Enemy Units&Level of Difficulty\\
\hline
2s3z&\makecell[c]{2 Stalkers\\3 Zealots}&\makecell[c]{2 Stalkers\\3 Zealots}&Easy\\
\hline
3s5z&\makecell[c]{3 Stalkers\\5 Zealots}&\makecell[c]{3 Stalkers\\5 Zealots}&Easy\\
\hline
1c3s5z&\makecell[c]{1 Colossus\\3 Stalkers\\5 Zealots}&\makecell[c]{1 Colossus\\3 Stalkers\\5 Zealots}&Easy\\
\hline
8m\_vs\_9m&8 Marines&9 Marines&Hard\\
\hline 
MMM2&\makecell[c]{1 Medivac\\2 Marauders\\7 Marines}&\makecell[c]{1 Medivac\\3 Marauders\\8 Marines}&Super Hard\\
\hline
bane\_vs\_bane&\makecell[c]{4 Banelings\\20 Zerglings}&\makecell[c]{4 Banelings\\20 Zerglings}&Hard\\
\hline
25m&25 Marines&25 Marines&Hard\\
\hline
27m\_vs\_30m&27 Marines&30 Marines&Super Hard\\
\hline
\end{tabular}}
\label{tab:challenges}
\normalsize
\end{table}

\subsection{Validation}
\begin{figure}[h]
    \centering
    \subfigure[1c3s5z]{\includegraphics[width=0.32\textwidth]{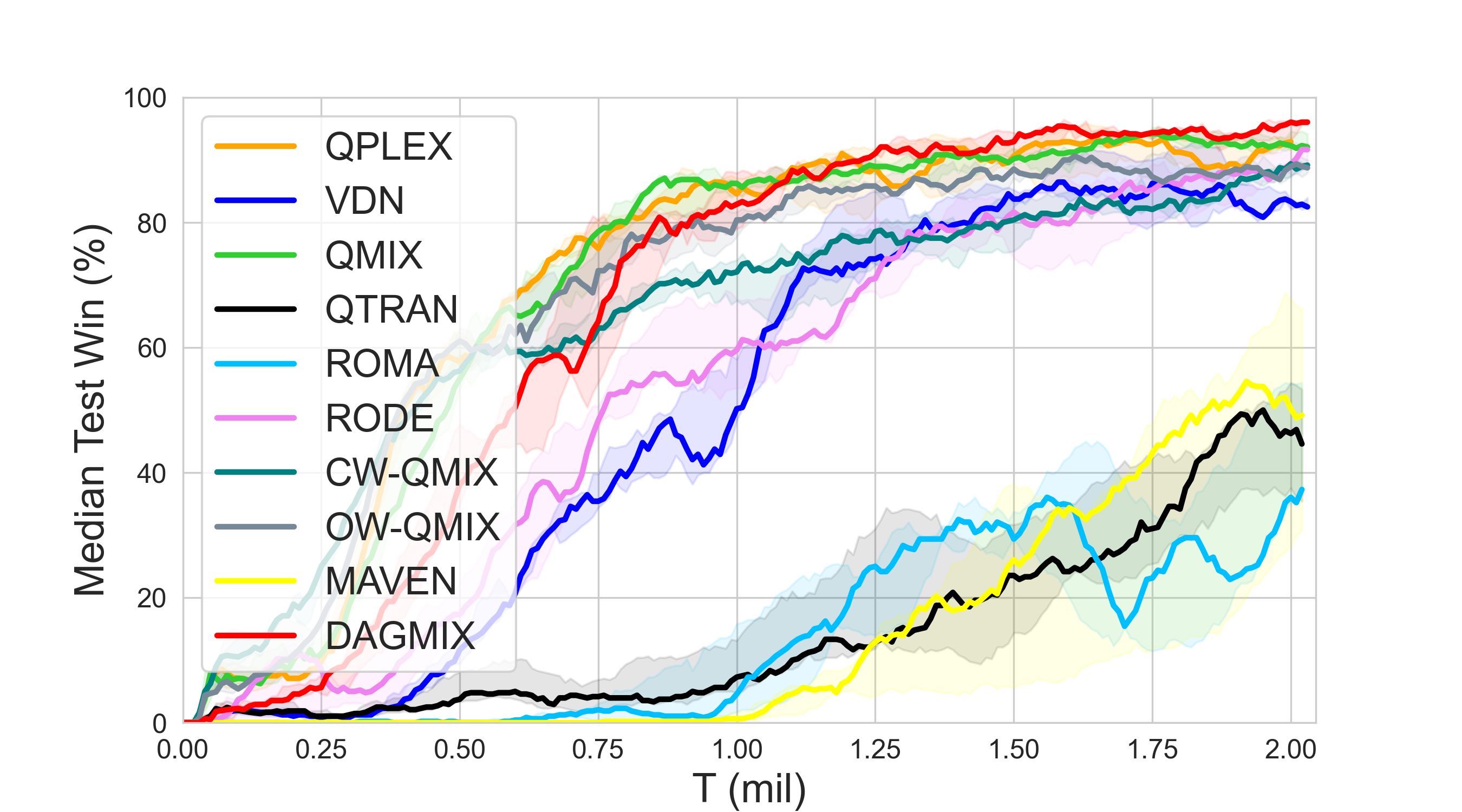}\label{fig:1c3s5z}}
    \subfigure[8m\_vs\_9m]{\includegraphics[width=0.32\textwidth]{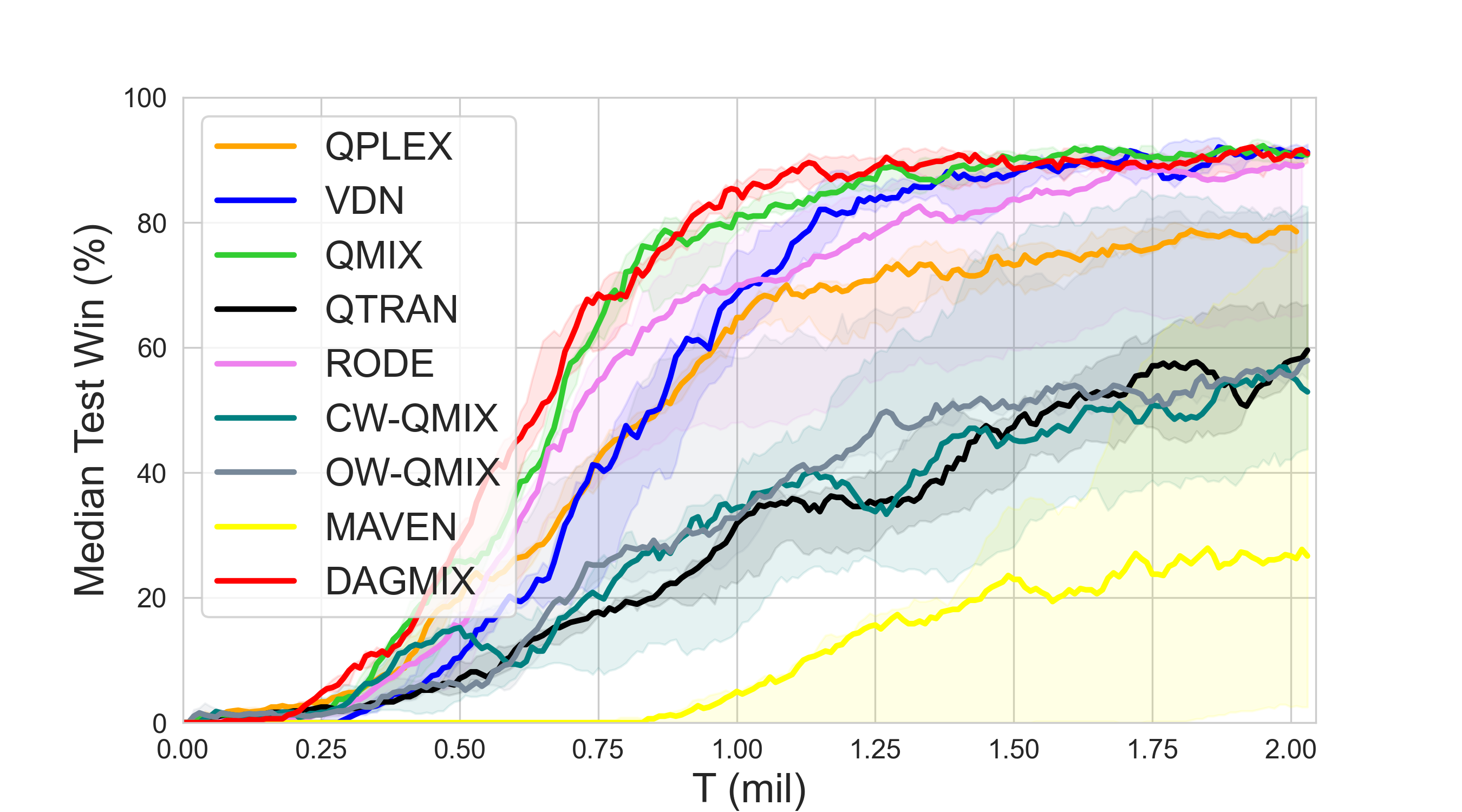}\label{fig:8m_vs_9m}}
    \subfigure[MMM2]{\includegraphics[width=0.32\textwidth]{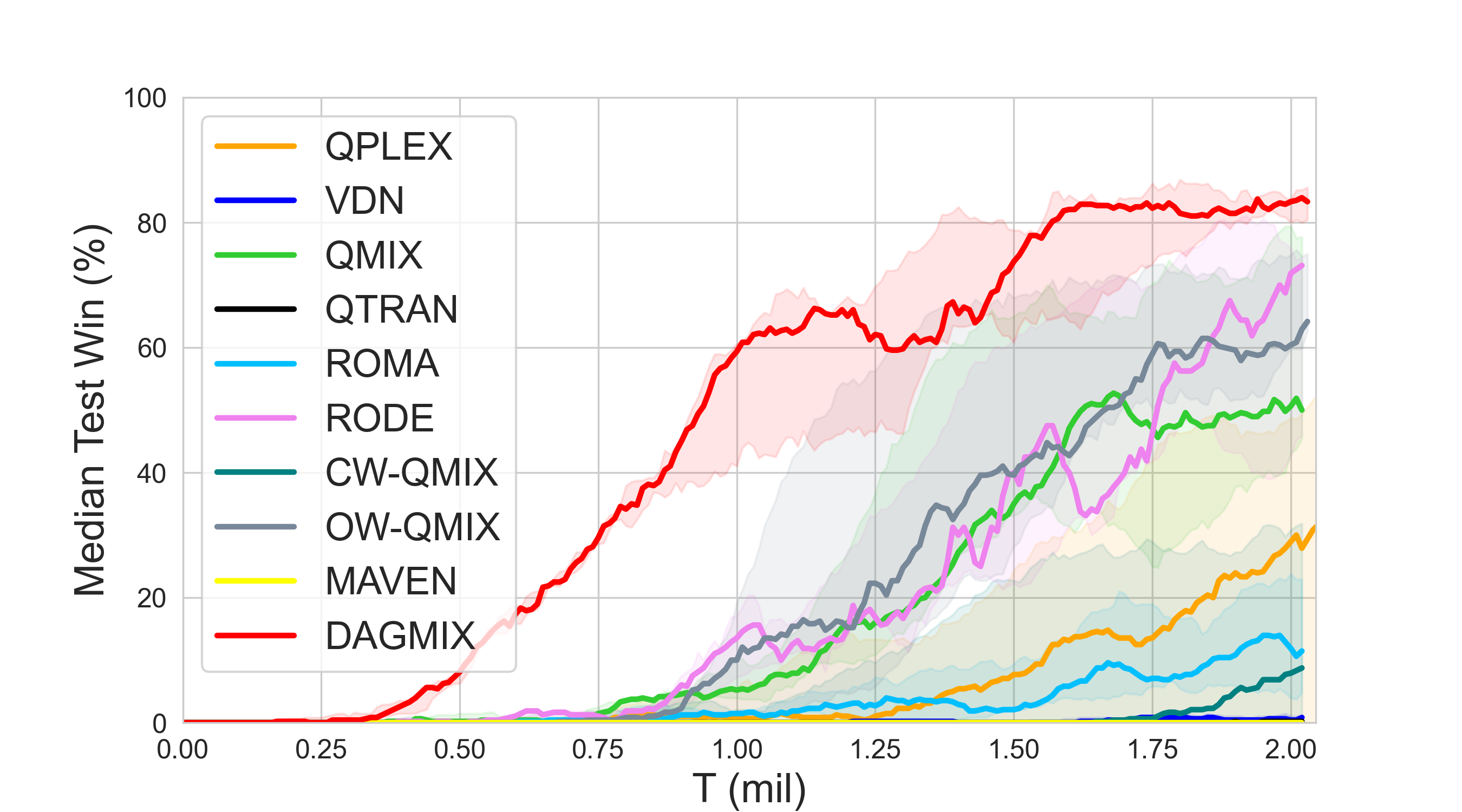}\label{fig:MMM2}}
    \subfigure[bane\_vs\_bane]{\includegraphics[width=0.32\textwidth]{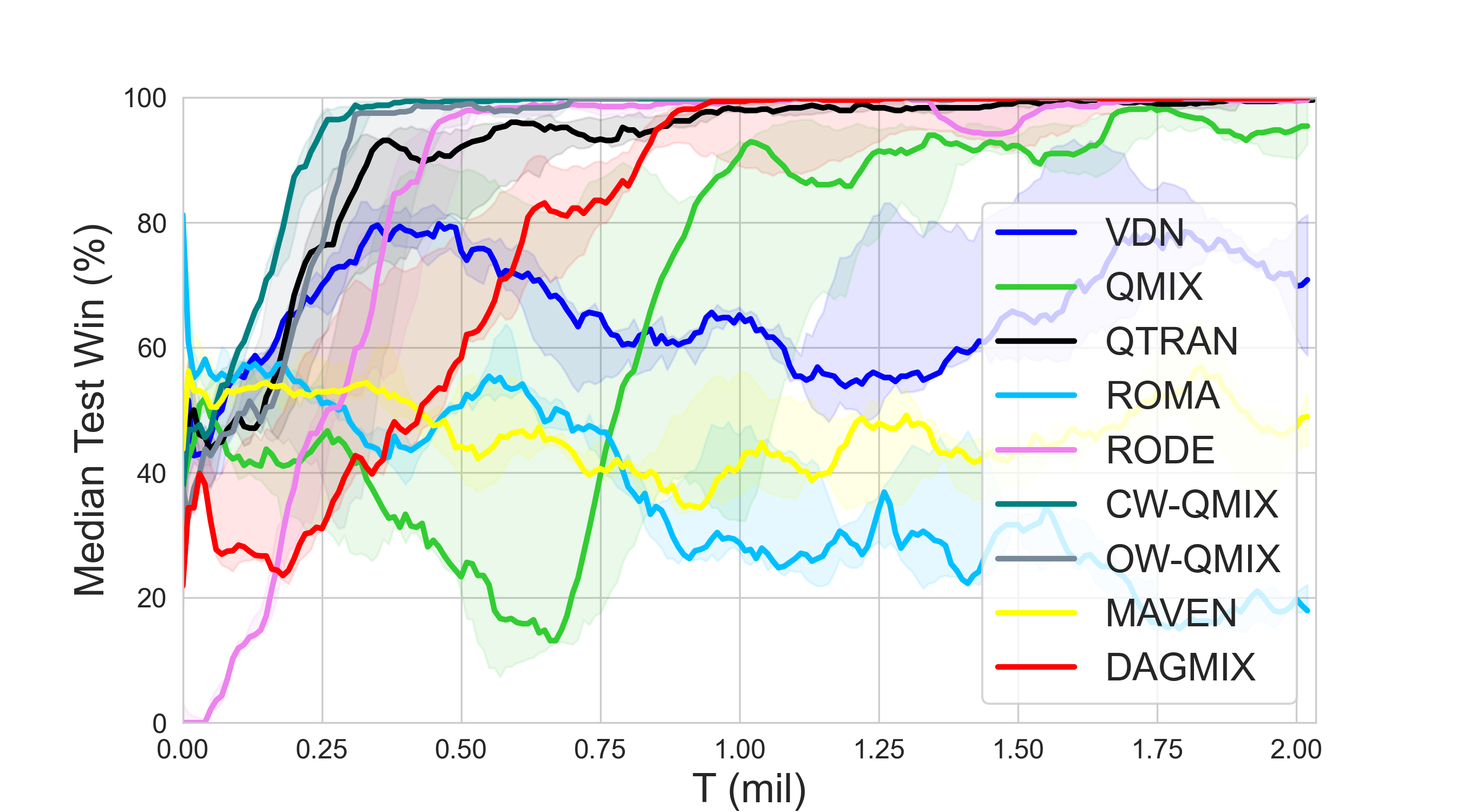}\label{fig:bvb}}
    \subfigure[25m]{\includegraphics[width=0.32\textwidth]{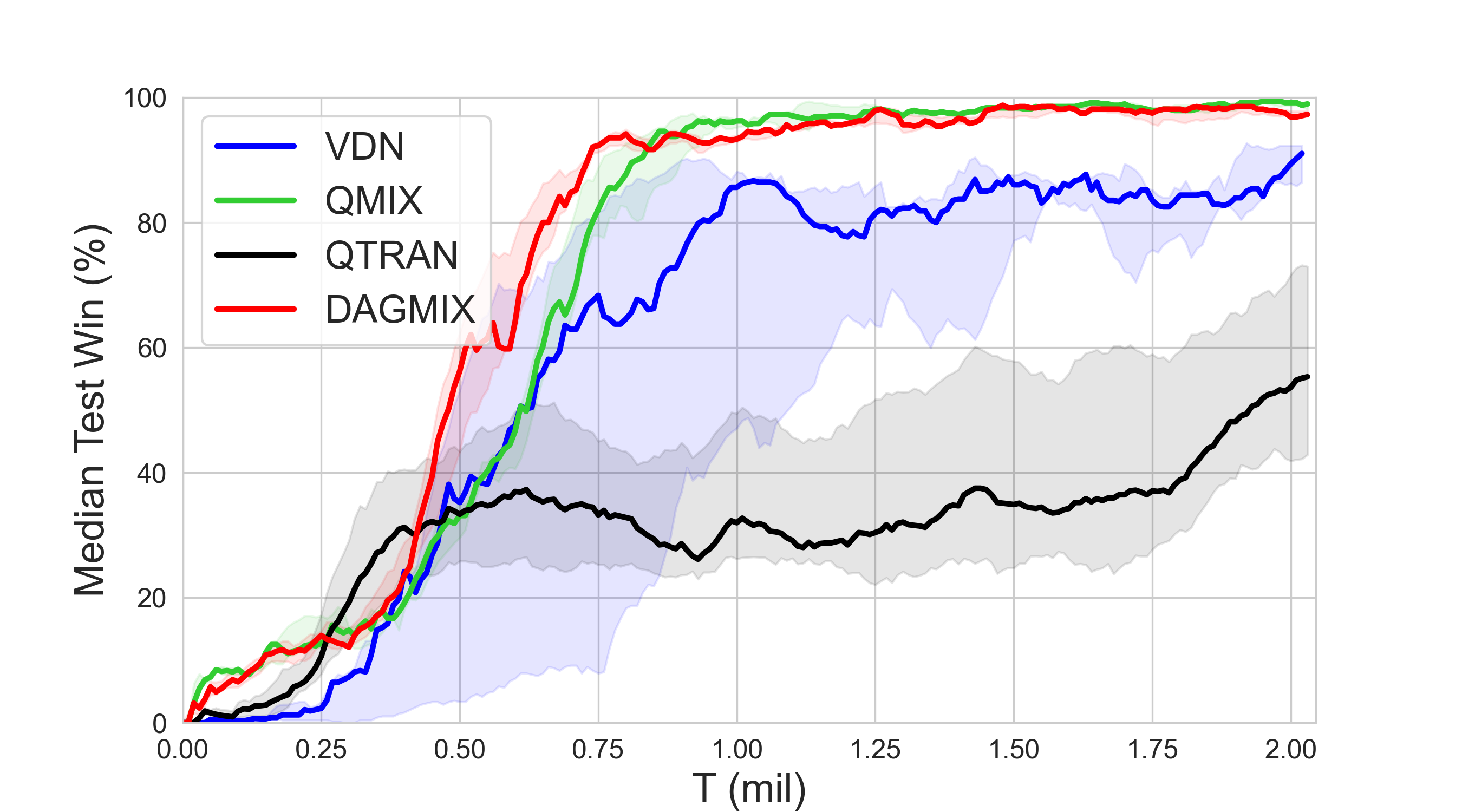}\label{fig:25m}}
    \subfigure[27m\_vs\_30m]{\includegraphics[width=0.32\textwidth]{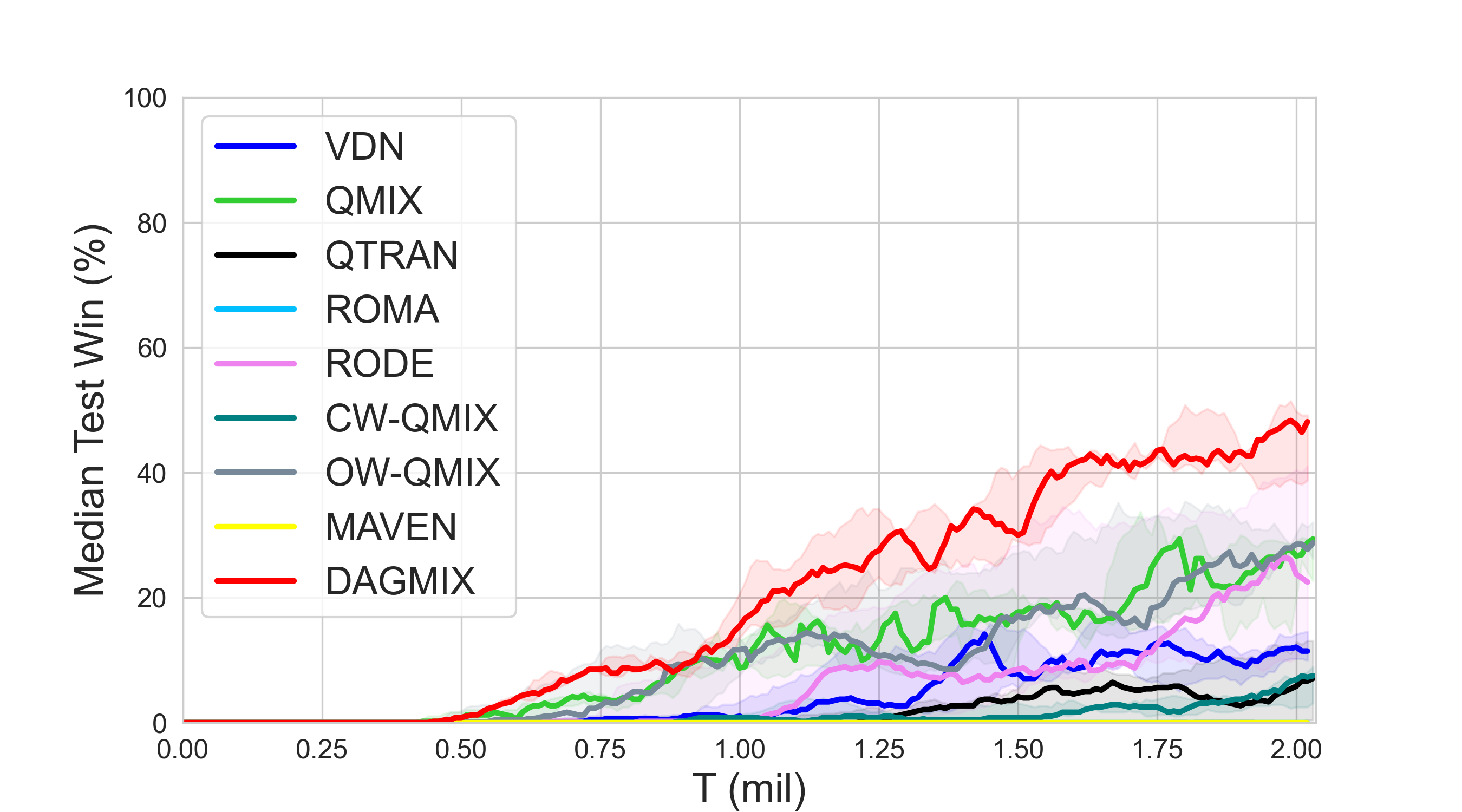}\label{fig:27m_vs_30m}}
    \caption{Overall results in different challenges.}
    \label{fig:results}
\end{figure}

On every challenge, we have run each algorithm 2 million steps for 5 times with different random seeds and recorded the changes in win ratio. To reduce the contingency in validation, we take the average win ratio of the recent 15 tests as the current results. The performances of DAGMIX and the baselines are shown in Figure~\ref{fig:results}, where the solid line represents the median win ratio of the five experiments using the corresponding algorithm, and the 25-75\% percentiles of the win ratios are shaded. Detailed win ratio data is displayed in Table~\ref{tab:Results}.

\begin{table}[h]
\small
\centering
\caption{Median performance and of the test win ratio in different scenarios.}
\resizebox{0.8\linewidth}{!}{
\begin{tabular}{c|cccc}
\ &DAGMIX&VDN&QMIX&QTRAN\\
\hline
1c3s5z&93.75&\textbf{96.86}&94.17&11.67\\
8m\_vs\_9m&92.71&\textbf{94.17}&\textbf{94.17}&70.63\\
MMM2&\textbf{86.04}&4.79&45.42&0.63\\
bane\_vs\_bane&\textbf{100}&94.79&99.79&99.58\\
25m&\textbf{99.79}&87.08&\textbf{99.79}&35.20\\
27m\_vs\_30m&\textbf{49.58}&9.58&33.33&18.33\\
\hline
\end{tabular}}
\label{tab:Results}
\normalsize
\end{table}
Among these baselines, QMIX is recognized as the SOTA method and the main competitor of DAGMIX due to its stability and effectiveness on most tasks. It can be clearly observed that the performance of DAGMIX is very close to the best baseline in relatively easy challenges such as \emph{1c3s5z} and \emph{8m\_vs\_9m}. As the scenarios become more asymmetric and complex, DAGMIX begins to show its strengths and exceed the performances of the baselines. In \emph{bane\_vs\_bane} and \emph{25m}, the win ratio curve of DAGMIX lies on the left of QMIX, indicating a faster convergence rate. \emph{MMM2} and \emph{27m\_vs\_30m} are classified as super hard scenarios on which most algorithms perform very poorly, due to the huge amount of agents and significant disparity between the opponent's troops and ours. However, Figure~\ref{fig:MMM2}~and~\ref{fig:27m_vs_30m} show that our method makes tremendous progress on the performance, as DAGMIX achieves unprecedented win ratios on these two tasks. Notably, the variance of DAGMIX's training results is relatively small, indicating its robustness against random perturbations.

\subsection{Ablation}
We conducted ablation experiments to demonstrate that the dynamic graph generated by the hard-attention mechanism is key to the success of DAGMIX. First, while many studies have improved original models by replacing the fully connected layer with the attention layer, we doubt if DAGMIX also benefits from such tricks, so we select Qatten~\cite{yang2020qatten} as the control group which also aggregates individual Q-values through attention mechanism but lacks a graph structure. Second, to confirm the effectiveness of the sparse dynamic graph over a fully connected one, we set all values in the adjacency matrix $\mathbf{A}$ to 1 which is referred to as \textit{Fully Connected Graph MIX} (FCGMIX) for comparison.

\begin{figure}[t]
    \centering
    \subfigure[2s3z]{\includegraphics[width=0.4\textwidth]{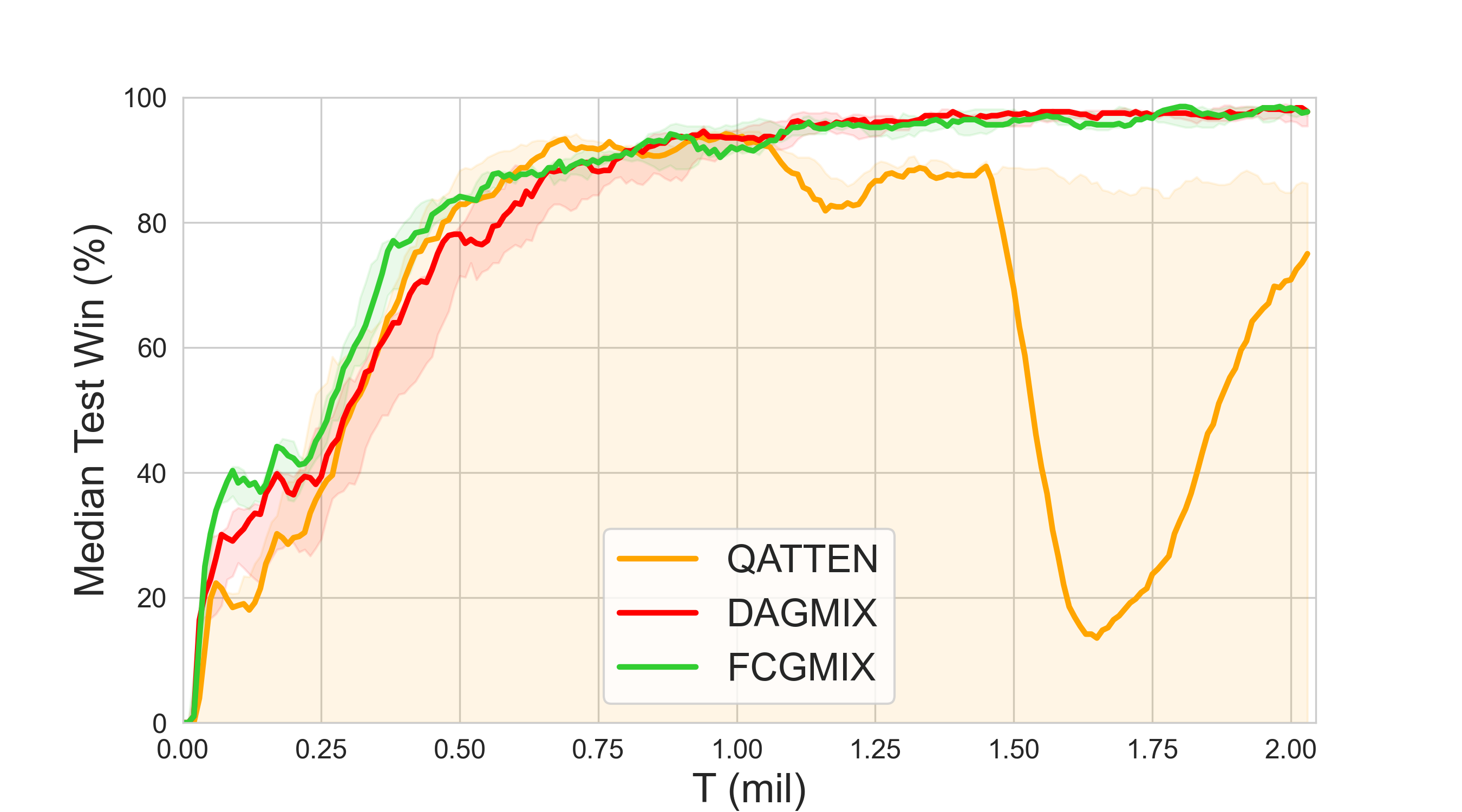}\label{fig:2s3z_ablation}}
    \subfigure[8m\_vs\_9m]{\includegraphics[width=0.4\textwidth]{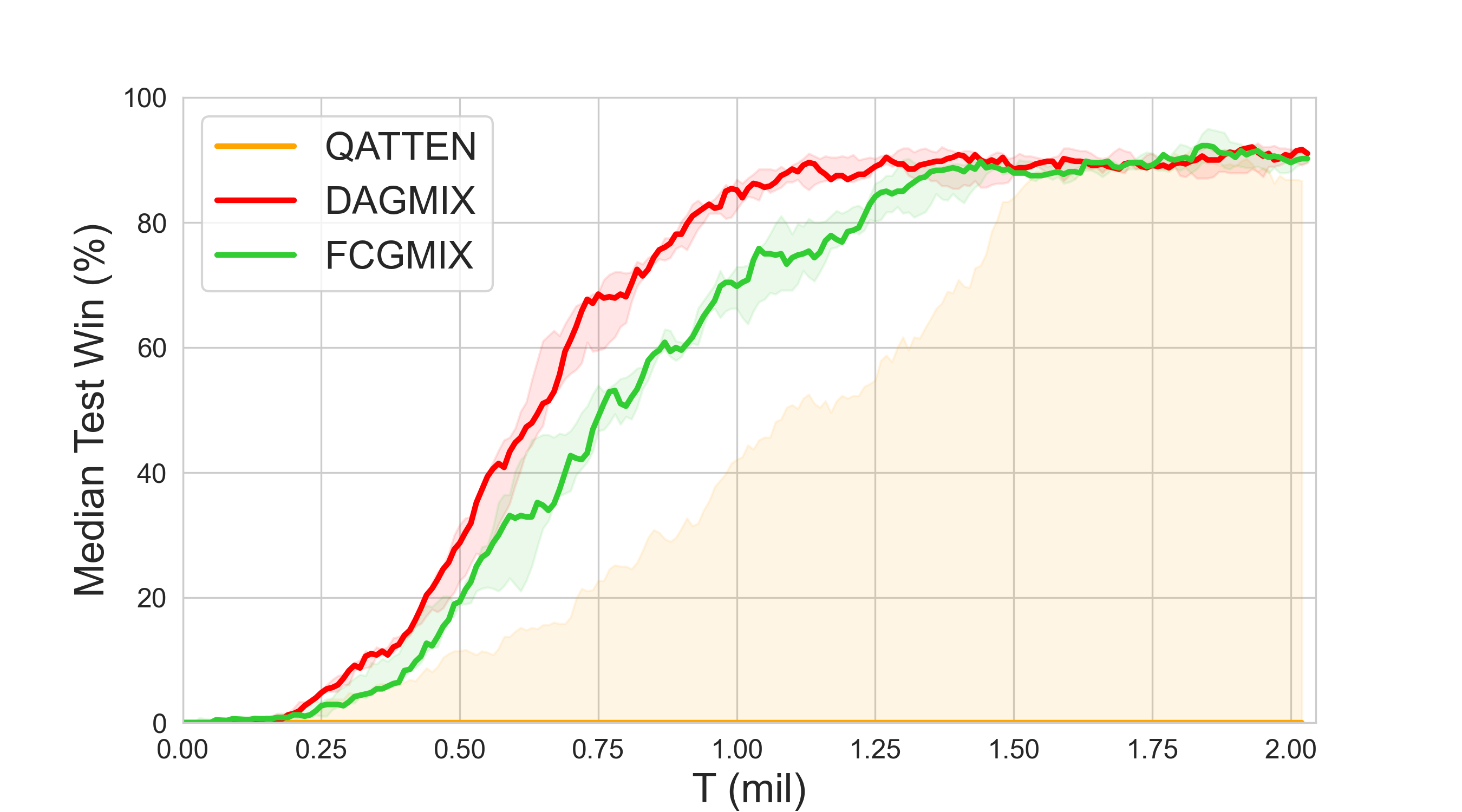}\label{fig:8m_vs_9m_ablation}}
    \subfigure[MMM2]{\includegraphics[width=0.4\textwidth]{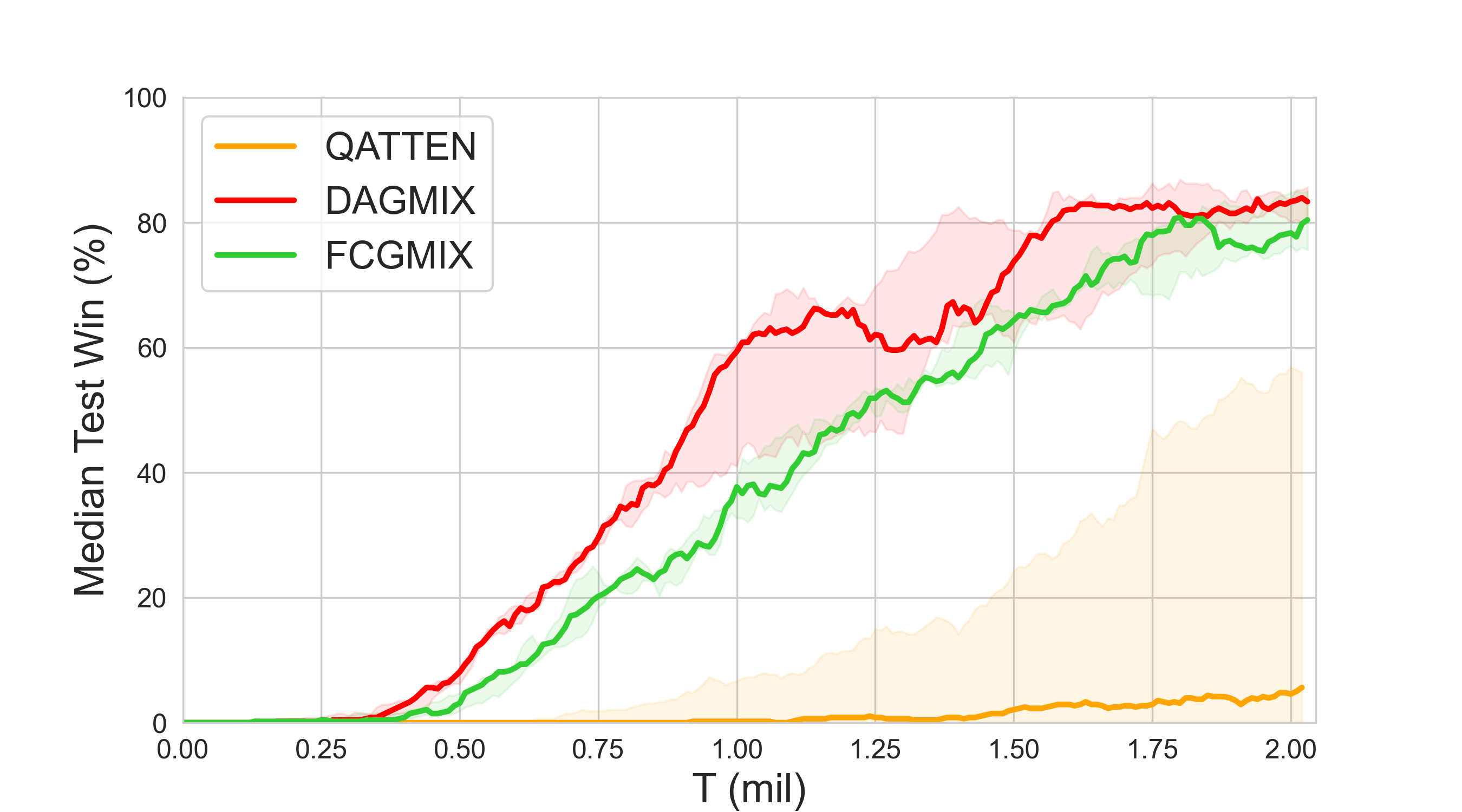}\label{fig:MMM2_ablation}}
    \subfigure[27m\_vs\_30m]{\includegraphics[width=0.4\textwidth]{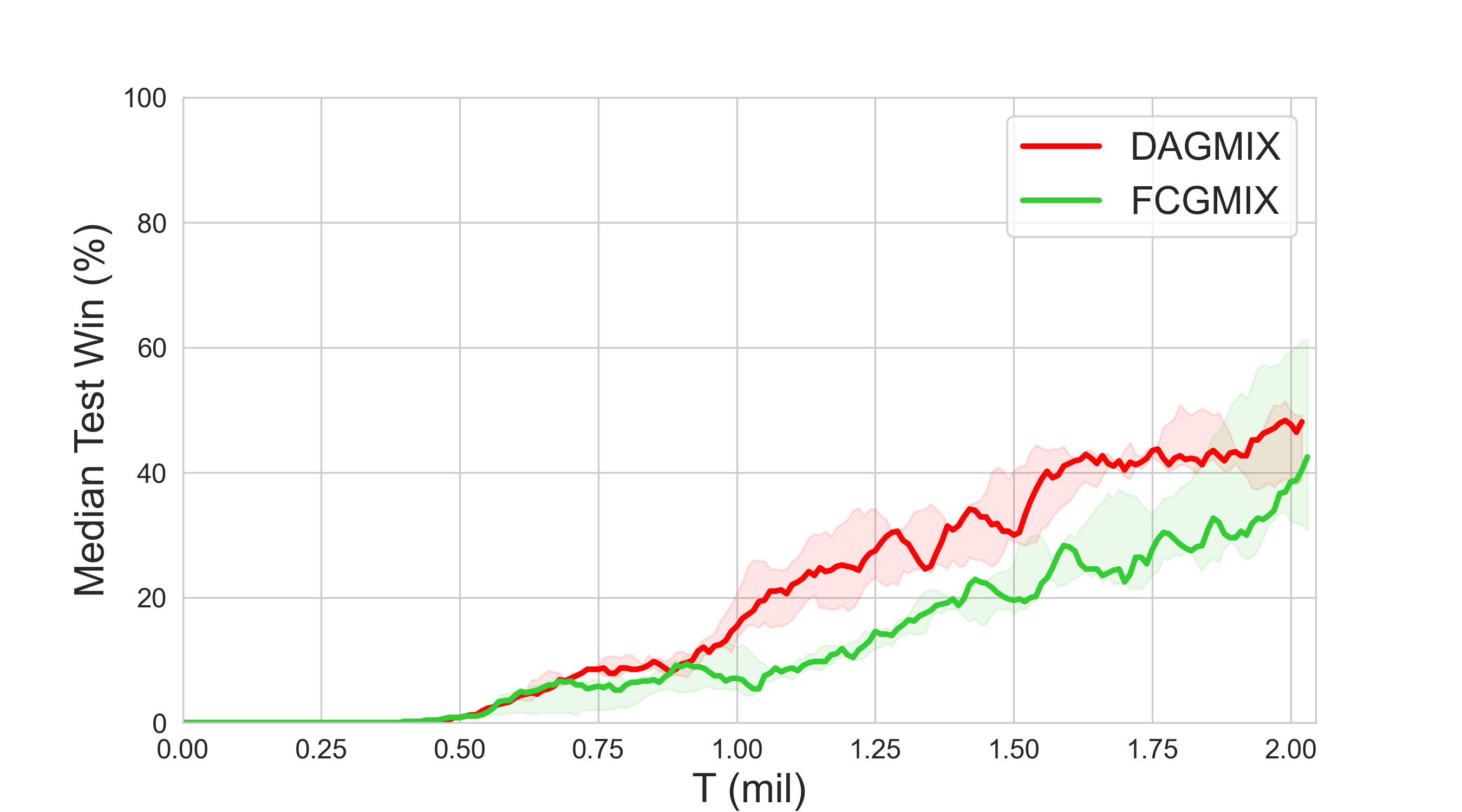}\label{fig:27m_vs_30m_ablation}}
    \caption{Ablation experiments results.}
    \label{fig:ablation}
\end{figure}

From Figure~\ref{fig:ablation} we find Qatten gets a disastrous win ratio with high variance, indicating that simply adding an attention layer does not necessarily improve performance. When there are fewer agents, the fully connected graph is more comprehensive and does not have much higher complexity than the dynamic graph, so in \emph{2s3z} FCGMIX performs even slightly better than DAGMIX. However, as the number of agents increases and the complete graph becomes more complex, DAGMIX is gaining a clear advantage thanks to the optimized graph structure.

\begin{figure}[h]
    \centering
    \subfigure[3s5z]{\includegraphics[width=0.4\textwidth]{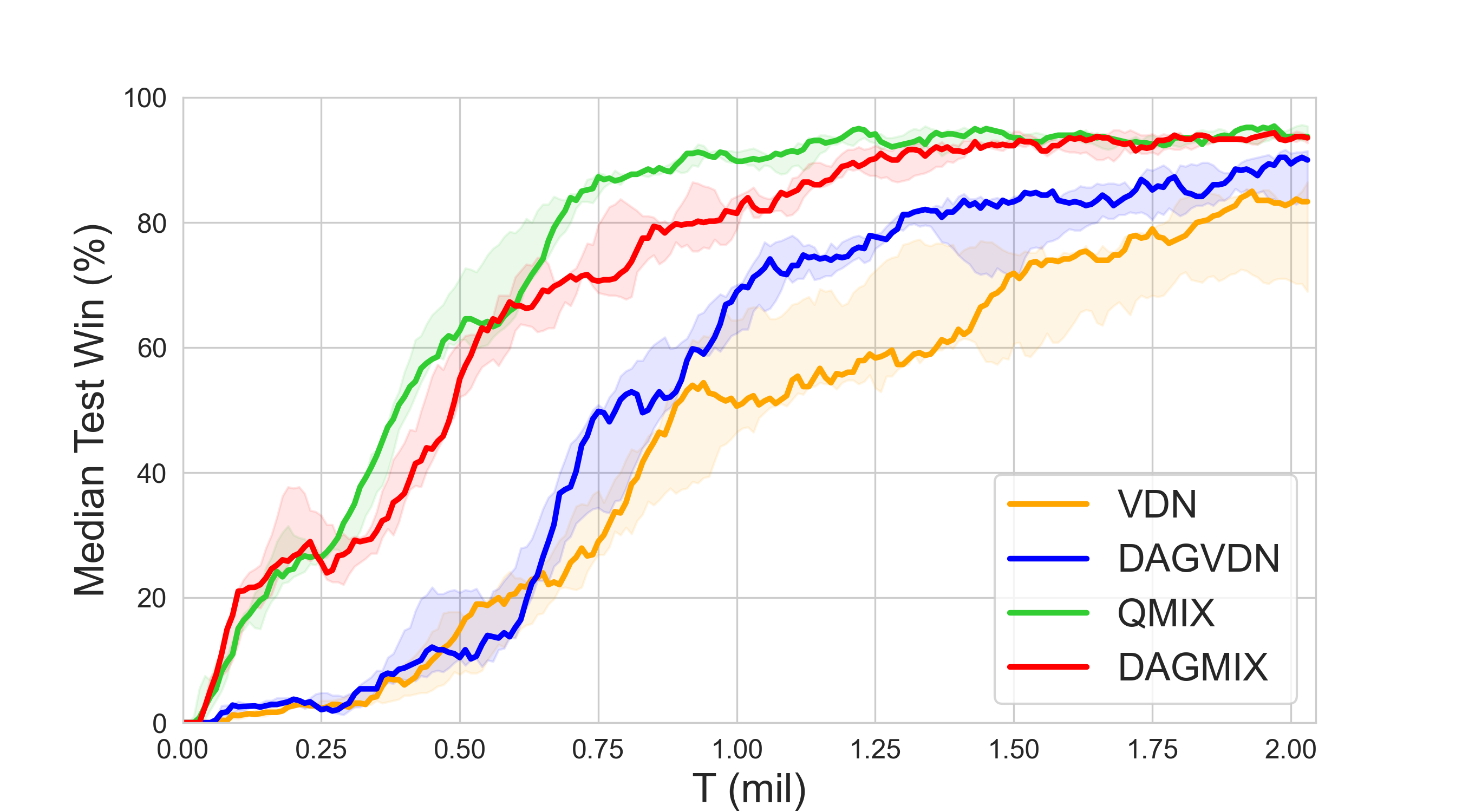}\label{fig:3s5z_vdn}}
    \subfigure[1c3s5z]{\includegraphics[width=0.4\textwidth]{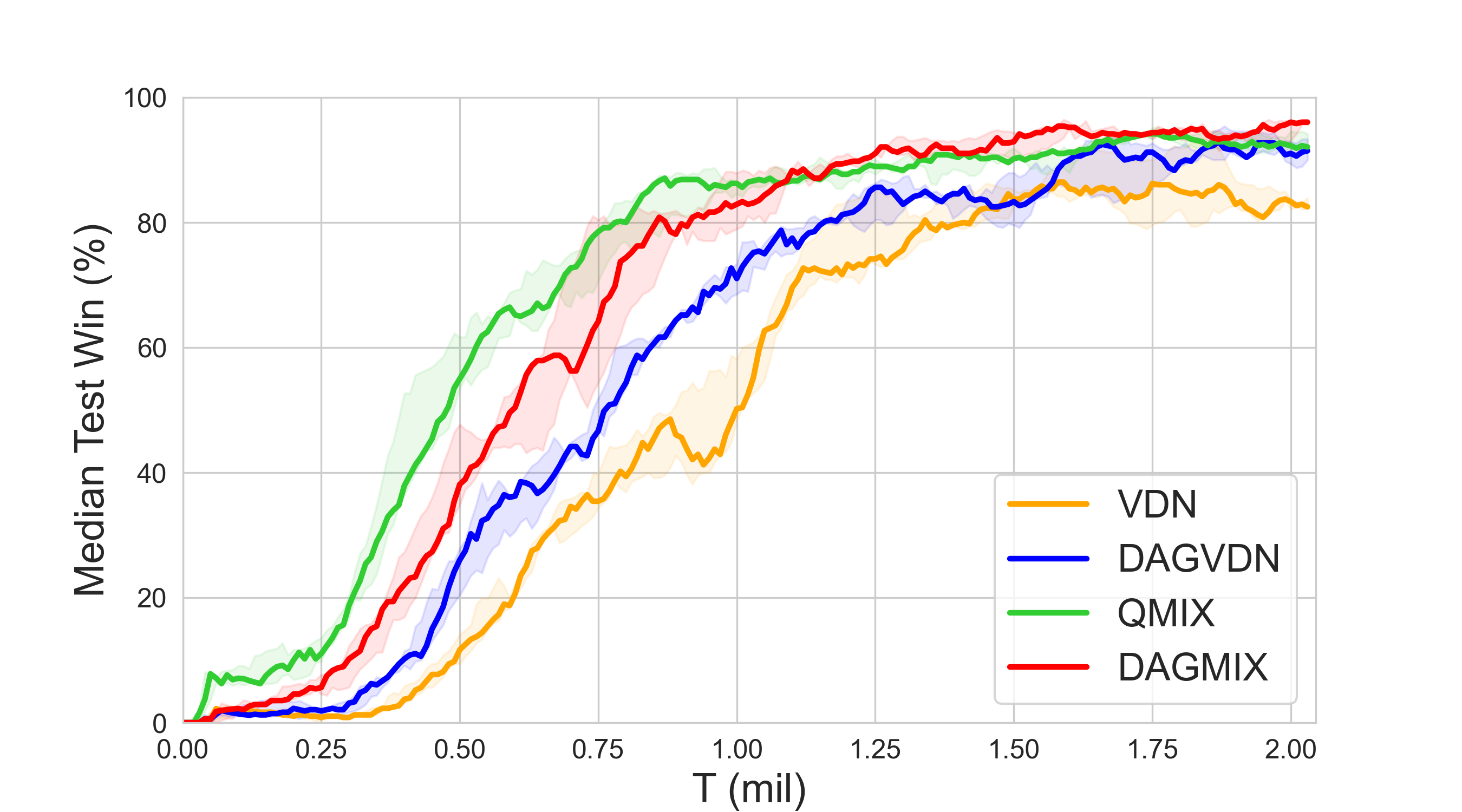}\label{fig:1c3s5z_vdn}}
    \subfigure[8m\_vs\_9m]{\includegraphics[width=0.4\textwidth]{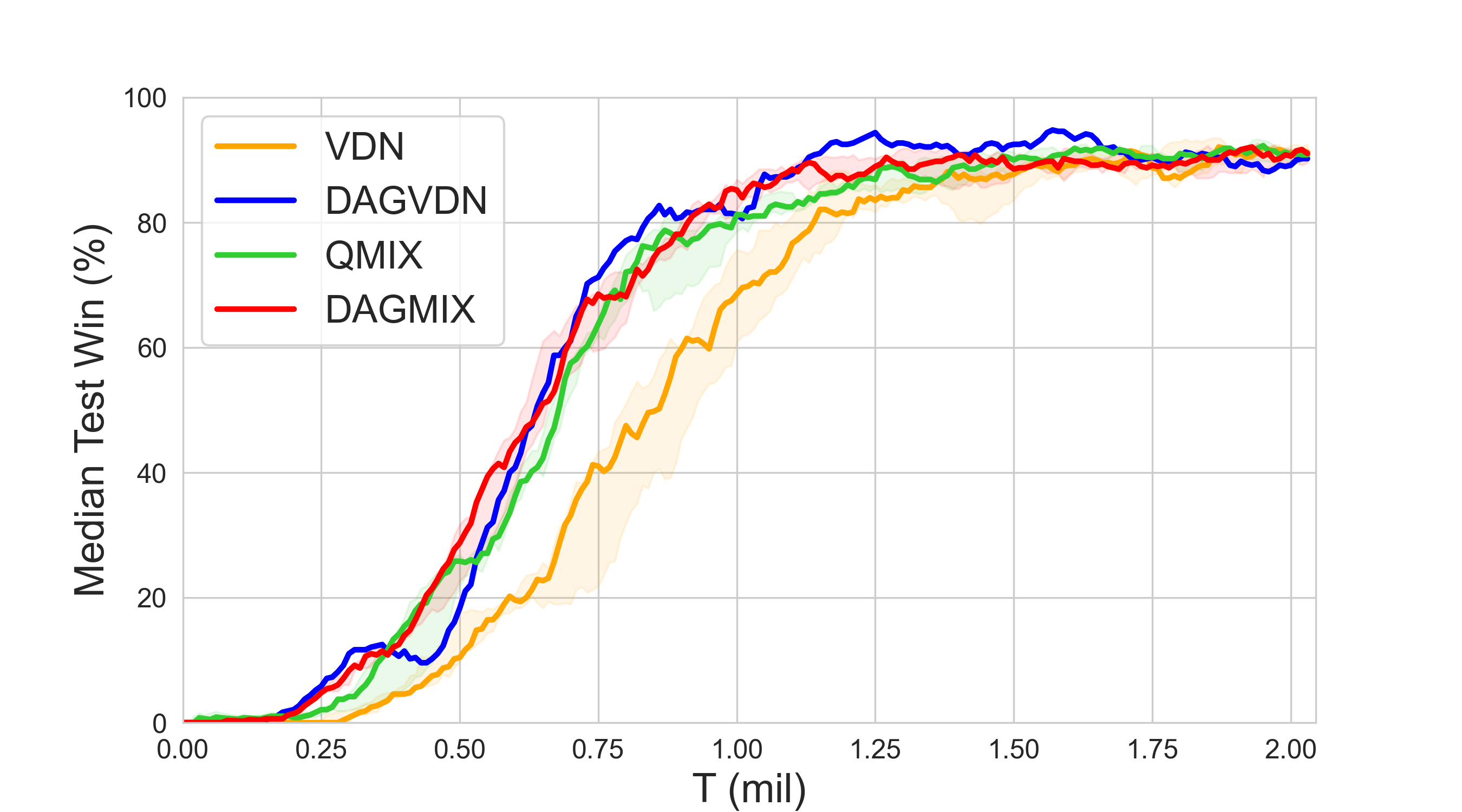}\label{fig:8m_vs_9m_vdn}}
    \subfigure[MMM2]{\includegraphics[width=0.4\textwidth]{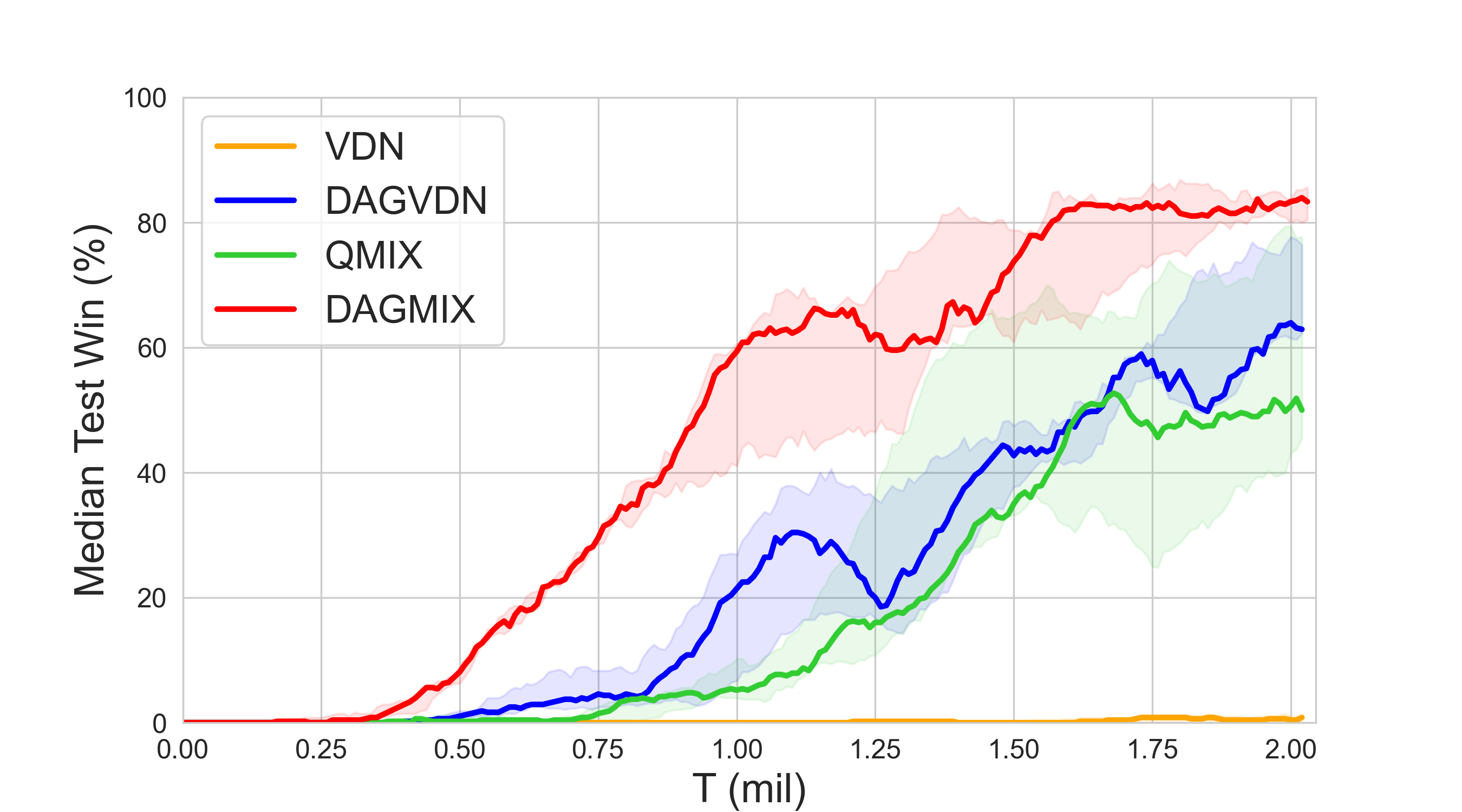}\label{fig:MMM2_vdn}}
    \caption{The results of DAGMIX and DAGVDN compared to original algorithms.}
    \label{fig:DAGVDN}
\end{figure}

It was noted in Section 3.2 that DAGMIX is a framework that enhances the base algorithm on large-scale problems. Here we combine VDN with our dynamic graph, denoted as DAGVDN, to investigate the improvements upon VDN, as well as DAGMIX upon QMIX. The results are presented in Figure~\ref{fig:DAGVDN}. Similar to DAGMIX, DAGVDN outperforms VDN slightly in relatively easy tasks, while it shows significant superiority in more complex scenarios that VDN couldn't handle.

\section{Conclusion}
In this paper, we propose DAGMIX, a cooperative MARL algorithm based on value factorization. It combines the individual Q-values of all agents through operations on a real-time generated dynamic graph and provides a more interpretable and precise estimation of the global value to guide the training process. DAGMIX catches the intrinsic relationship between agents and allows end-to-end learning of decentralized policies in a centralized manner. 

Experiments on SMAC show the prominent superiority of DAGMIX when dealing with large-scale and asymmetric problems. In other scenarios, DAGMIX still demonstrates very stable performance which is comparable to the best baselines. We believe DAGMIX provides a reliable solution to multi-agent coordination tasks.

There are still points of improvement for DAGMIX. First, it seems practical to apply multi-head attention in later implementation. Besides, the dynamic graph in DAGMIX is not always superior to the complete graph when dealing with problems with fewer agents. On the basis of the promising findings presented in this paper, work on the remaining issues is continuing and will be presented in future papers.

\subsubsection*{Acknowledgements}
This work was supported by the Strategic Priority Research Program of the Chinese Academy of Science, Grant No.XDA27050100.

%
%
%

\bibliographystyle{splncs04}
\bibliography{mybibliography}

\end{document}